\documentclass{article}
\usepackage[T1]{fontenc} 
\usepackage[utf8]{inputenc} 
\usepackage{ismir,amsmath,cite,url}
\usepackage{graphicx}
\usepackage{color}
\usepackage{booktabs}
\usepackage{tabularray}
\usepackage{multirow}
\usepackage{amsmath}
\usepackage{amssymb}
\usepackage{enumitem}
\usepackage{makecell}
\usepackage{lineno}
\usepackage{xcolor}
\usepackage{array}
\usepackage{float}
\usepackage{colortbl}
\usepackage{microtype}
\usepackage{siunitx}

\DeclareSIUnit \bpm {bpm}

\title{Symbolic Music Representations for Classification Tasks: A Systematic Evaluation}

\multauthor
{Huan Zhang$^1$ \hspace{1cm} Emmanouil Karystinaios$^2$ \hspace{1cm} Simon Dixon$^1$} { \bfseries{Gerhard Widmer$^2$ \hspace{1cm} Carlos Eduardo Cancino-Chacón$^2$ \hspace{1cm}}\\
 $^1$ Queen Mary University of London, United Kingdom\\
$^2$ Johannes Kepler University, Austria\\
}

\def\authorname{H. Zhang, E. Karystinatos, S. Dixon, G. Widmer, C.E. Cancino-Chacón}

\begin{document}

\maketitle
\begin{abstract}

Music Information Retrieval (MIR) has seen a recent surge in deep learning-based approaches, which often involve encoding symbolic music (i.e., music represented in terms of discrete note events) in an image-like or language-like fashion. However, symbolic music is neither an image nor a sentence, and research in the symbolic domain lacks a comprehensive overview of the different available representations. In this paper, we investigate matrix (piano roll), sequence, and graph representations and their corresponding neural architectures, in combination with symbolic scores and performances on three piece-level classification tasks. We also introduce a novel graph representation for symbolic performances and explore the capability of graph representations in global classification tasks. Our systematic evaluation shows advantages and limitations of each input representation. Our results suggest that the graph representation, as the newest and least explored among the three approaches, exhibits promising performance, while being more light-weight in training.



\end{abstract}
\section{Introduction}\label{sec:introduction}

The deep learning boom has profoundly impacted MIR, including research involving
symbolic music representations (MIDI, scores, etc.).
A large body of recent literature focuses on adapting existing architectures from computer vision and natural language processing to the field of symbolic MIR. These approaches often treat music data as an image (piano roll), as a sequence of language tokens, or, more recently, as a graph. However, a piece of music is neither an image nor a sentence or graph, therefore, a critical question still remains open concerning the choice of input representations for symbolic music.

A source of complexity in symbolic music arises from the different modalities of data such as scores and performances. A score contains information about music notation and often includes rich hierarchically structured information such as metrical structure and voicing. Symbolic music performances, on the other hand, such as those recorded on a MIDI-capable instrument, consist of a stream of controller events. Extracting a hierarchical structure from such a stream is not a trivial task~\cite{liu2022performance,Temperley:2009,Temperley:CBMS2004}. Furthermore, such performance data omit some of the rich information that a score provides, such as pitch spelling and articulation markings, but instead, it can include information about expression, timing, local tempo, and performance dynamics.

Recent research has produced relatively large datasets containing scores and performances at the symbolic level, including efforts to align these~\cite{Zhang2022ATEPPPerformance, asap, foscarin2022match}. Motivated by these developments, we present an attempt to shed light on questions revolving around the input representation of symbolic music for deep-learning-based MIR. We formulate an empirical framework where we test multiple input representations, models, and piece-level classification tasks. 

In terms of input representations, we investigate piano rolls, tokenized sequences, and graphs. We evaluate multiple models based on these representations on three different tasks: composer classification, performer classification, and (playing) difficulty assessment. Furthermore, having datasets containing both performances and their corresponding scores such as ATEPP and ASAP~\cite{Zhang2022ATEPPPerformance, asap}, allows us to apply each combination of representation and task to either score or performance. 
Our goal is to contribute an experimental overview of different symbolic music representations.
The contributions of this work are threefold:
\begin{enumerate}[noitemsep]
    \item We investigate the performance and complexity of matrix, sequence and graph input representations, and their corresponding neural architectures (respectively Convolutional Neural Networks, Transformers, and Graph Neural Networks). 
    \item  We compare the impact that the different information contained in symbolic scores and performances has on different piece-level classification tasks.
    \item We introduce a new graph representation for symbolic performances, and explore the capability of graph representations in  classification tasks.
\end{enumerate}
\section{Related Work}\label{sec:related_work}

The complexity of representing music data has been discussed in the literature \cite{Xenakis1992FormalizedComposition, Harris1991RepresentingSymbolically,babbit:1965}. Wiggins et al.\ \cite{Wiggins1993FrameworkSystems} analyzed the trade-offs of music representation systems with respect to expressive completeness and structural generality.  In the age of deep learning, such considerations are still relevant regarding the variety of machine-readable representations such as piano rolls, MIDI-like sequences, NoteTuples, and Musical Spaces \cite{Walder2016ModellingRoll, Prang2021RepresentationMusic}. In this section, we focus on three symbolic representations (matrix, sequence, and graphs) and discuss their respective strengths and limitations. 

\begin{figure*}[t]
    \centering
    \includegraphics[width=\linewidth]{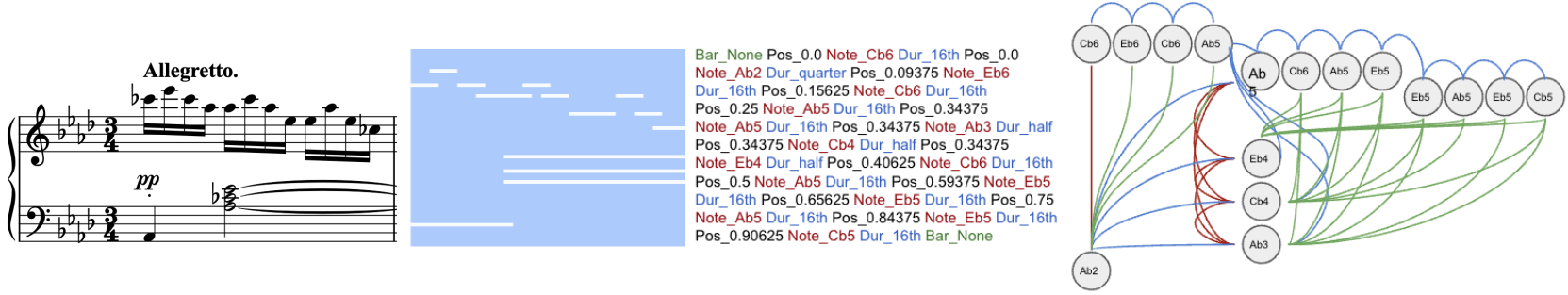}
    \caption{Excerpt of Schubert's \textit{Impromptu Op. 90 No.4} and its input visualizations (from left to right): generic matrix, sequence (REMI-like) and graph.}
    \label{fig:representations}
\end{figure*}

\vspace{-2.5ex}
\paragraph*{Music as a Matrix:}\label{subsec:label_matrix}

Similar to audio spectrograms, a pitch-time representation that is typically used as input to a CNN, the piano roll representation of music naturally emerges as the symbolic equivalent. Piano rolls have been widely applied in tasks such as automatic music transcription \cite{Benetos2012Score-informedTutoring, Kong2021High-ResolutionTimes}, classification of piece-level attributes such as difficulty and composer \cite{Ghatas2022HybridMusic, Kim2020DeepRepresentation, Velarde2016ComposerPiano-rolls, Foscarin2022ConceptClassifier}, as well as generation of music accompaniment or performed dynamics \cite{Dong2018MuseGANAccompaniment, VanHerwaarden2014PredictingNetworks}. 

A piano roll is a bare-bones representation of symbolic music data, and, therefore, information such as key signatures, articulation annotations, metrical structure, different instrument parts, and voicing structure are not encoded in the representation\cite{Briot2017DeepSurvey, Walder2016ModellingRoll}.

\vspace{-2.5ex}
\paragraph*{Music as a Sequence:}\label{subsec:tokens}

Modeling symbolic music as sequences has a longstanding tradition in MIR. 
The multiple viewpoint system is a sequence representation that has been widely used for music analysis, generation, and classification~\cite{ConklinWitten:1995,ConklinClassification:2013,WhorleyConklin:2016,Conklin:2016}, as well as the basis for cognitively plausible models of expectation \cite{Pearce:2018,Pearce:idyom}.
In this system, musical elements are represented by viewpoints \cite{Conklin1995MultiplePrediction}, which are abstract functions mapping musical events to abstract derived features like pitch, interval, and melodic contour. 

With the advances of deep learning-based language models, sequential representation of music as \textit{language tokens} has recently received a lot of attention in sequence-to-sequence generative tasks from automatic orchestration \cite{Liu2022SymphonyModel} to description-based medley generation \cite{Rutte2022FigaroControl}. Similar to a stream of MIDI messages, various tokenization schemes encode music features such as pitch, onset time, duration, and velocity sequentially. Besides generation, large-scale pre-training using music sequences has been applied to downstream music understanding tasks \cite{Keller2021WhatLearn, Zeng2021MusicBERTPre-Training}. 

However, tokenized music sequence representations create difficulty for models to learn the dependency of long contexts. Length reduction methods such as Byte Pair Encoding (BPE) \cite{Liu2022SymphonyModel, Fradet2023ByteMusic} aim to address the length overflow problem by replacing the occurrence of frequent subsequences with new tokens. 


\vspace{-2.5ex}
\paragraph*{Music as a Graph:}\label{subsec:graphs}

A musical score can also be seen as a graph where notes form the vertices and relations between notes define the edges. 
Jeong and al.~\cite{jeong2019graph} introduced a graph modeling of a musical score for generating expressive performances. Recently, Karystinaios and Widmer~\cite{karystinaios2022cadence} presented a new modeling of the score graph based on three different note relations and a Graph Convolutional Network for cadence detection in classical music. 
A score graph can be homogeneous or heterogeneous, i.e. having one or several types of edges and/or vertices, respectively \cite{shi2016survey}. 
We will investigate both heterogeneous and homogeneous score graphs based on the representation used in \cite{karystinaios2022cadence}.

Graph Neural Networks have gained popularity in recent years, however, graph learning inherently presents some limitations, such as over-smoothing in deep graph networks~\cite{Li2019DeepGCNsCNNs} and restrictions of Message Passing, where information in graph neural networks flows only between edge relations predetermined by the representation (in contrast to a Transformer architecture where everything is interconnected~\cite{Velickovic2023EverythingNetworks}).

\section{Methodology}\label{sec:methodology}

In this section, we describe the methodology followed, the corpora used, and the experiments conducted to investigate in-depth the different symbolic representations. 

\subsection{Representation Design}
\label{sec:rep_design}

We briefly introduce a formal definition of each representation type, i.e.\ matrix, sequence, and graph. An example of the three representations is shown in Figure~\ref{fig:representations}.

\subsubsection{Matrix}

We define as a matrix representation of music a 2-dimensional array $\mathbf{M} \in \mathbb{N}^{H \times W}$ that depicts musical notes on the time axis, commonly referred to as a piano roll. The vertical axis consists of 128 possible values attributed to the MIDI pitch of note events, where we add three more optional fields for the \textit{una corda}, \textit{sostenuto}, and sustain pedals only applied on the MIDI performances.

In this work, we experimented with multiple channels as used in Onsets and Frames \cite{Hawthorne2018OnsetsTranscription}. The onset channel is a binarized roll with activations at onset timestamps, while the frame channel encodes the duration of the note and the velocity of the MIDI event. For scores, the velocity values are substituted by the voice index, i.e. the integer number assigned to a note to indicate the index among the number of independent voices.\footnote{
This voice information is commonly available in formats such as MusicXML, **Kern, and MEI. 
} 

\subsubsection{Sequence}

A symbolic music sequence $\mathbf{S} \in \mathbb{N}^{1 \times N}$ is defined by a series of discrete tokens that represent attributes of notes. Vocabularies such as $V_{\texttt{pitch}}, V_{\texttt{TimeShift}}, V_{\texttt{Vel}}$ assign semantic meanings to tokens, and different tokenization schemes translate into different grammars of sequence construction. In this work, we test three popular tokenization schemes: \textit{MIDILike} \cite{Oore2018ThisPerformance, Huang2018MusicTransformer}, \textit{REMI} \cite{Huang2020PopCompositions}, and \textit{CompoundWord} \cite{Hsiao2021CompoundHypergraphs} and use the implementation of the MidiTok library \cite{Fradet2021MiditokTokenization}.



As there is no existing tokenizer for processing scores, we implemented custom MusicXML tokenizers following MidiTok's framework, in the style of \textit{REMI} as well as \textit{CompoundWord}. The major difference is the timing of bars and event positions, as well as the addition of score-specific tokens such as $V_{\texttt{KeySig}}, V_{\texttt{Voice}}$.\footnote{Full documentation is provided with our open-source tokenizer in the project repository.} 

Byte Pair Encoding (BPE) is a tokenizer add-on technique that has recently been applied to music sequence learning~\cite{Fradet2023ByteMusic}. It consists of a data compression technique that replaces the most common token subsequences in a corpus with newly created tokens. BPE increases the vocabulary size and shortens the sequence length. We follow the best results from \cite{Fradet2023ByteMusic} and adopt a BPE with 4 times the original vocabulary size. On average, this reduced our sequence length between $55-65\%$ in both datasets.

\subsubsection{Graph} 

A homogeneous score graph $G$ is defined by a tuple $(V, E)$ of vertices and edges. $V$ is the set of notes in a musical score and $E \subseteq V \times V$.
Given a score with $N$ notes, we extract a matrix of $k$-dimensional note-wise features $X \in \mathbb{R}^{N\times k}$ based on features contained in the score or performance. A heterogenous score graph $G=(V, E, \mathcal{R})$ also includes a set of relation types $\mathcal{R}$ such that for every edge $e \in E$, $e$ is of type $r \in \mathcal{R}$ if a condition defined by $r$ holds. In our work, we consider the following relations between two notes $u, v$ which define the edges $e \in E$:
\begin{itemize}[noitemsep]
    \item $u$ and $v$ have the same onset, i.e.\ $\mathit{on}(v) = \mathit{on}(u)$, then $r = \textrm{onset}$;
    \item The offset of $u$ is the onset of $v$, i.e.\ $\mathit{off}(u) = \mathit{on}(v)$, then $r = \textrm{consecutive}$;
    \item The onset of $u$ lies between the onset of $v$ and the offset of $v$, i.e.\ $\mathit{on}(v) < \mathit{on}(u) \land \mathit{on}(u) < \mathit{off}(v)$, then $r = \textrm{overlap}$.
\end{itemize}

The above relations only hold in the case of score graphs. To adapt this to performance graphs, we use a window tolerance $t_{\textrm{tol}}$, such that if two notes $(u, v) \in E$ and:
\begin{itemize}[noitemsep]
    \item $|\mathit{on}(v) - \mathit{on}(u)| < t_{\textrm{tol}}$, then $r = \textrm{onset}$;
    \item $|\mathit{off}(u) - \mathit{on}(v)| < t_{\textrm{tol}}$, then $r = \textrm{consecutive}$;
    \item $\mathit{on}(v) < \mathit{on}(u) \land \mathit{on}(u) < \mathit{off}(v)$, then $r = \textrm{overlap}$.
\end{itemize}
In our configurations, for all graphs created from performance MIDI, we set $t_{\textrm{tol}} = \SI{30}{\milli\second}$, a perceptual threshold of expressive timing \cite{Goebl2001MelodyArtifact}. 
In addition to the above relations, we consider the possibility of adding an inversely directed edge for the overlap and the consecutive edge types, and we name the inclusion of such edges \textit{inverse edges}. 
For a homogeneous graph $G_{\textrm{hom}}$ and heterogeneous graph $G_{\textrm{het}}$, $e \in G_{\textrm{hom}} \implies e \in G_{\textrm{het}}$.

The node features $X$ are divided into two categories, the basic and the advanced features. The basic features are implicitly contained in any score or performance note such as one-hot encoding of pitch class and octave of the note's pitch, and duration information. The advanced features contains articulation, dynamics, and notation information from the \textit{Partitura} python package~\cite{partitura2022mec}. The detailed computation of these features can be found in original partitura paper \cite{Cancino-Chacon2022PartituraProcessing} and the basis mixer \cite{Cancino-Chacon2018ComputationalReview}.

\begin{figure}[t]
    \centering
    \includegraphics[width=\linewidth]{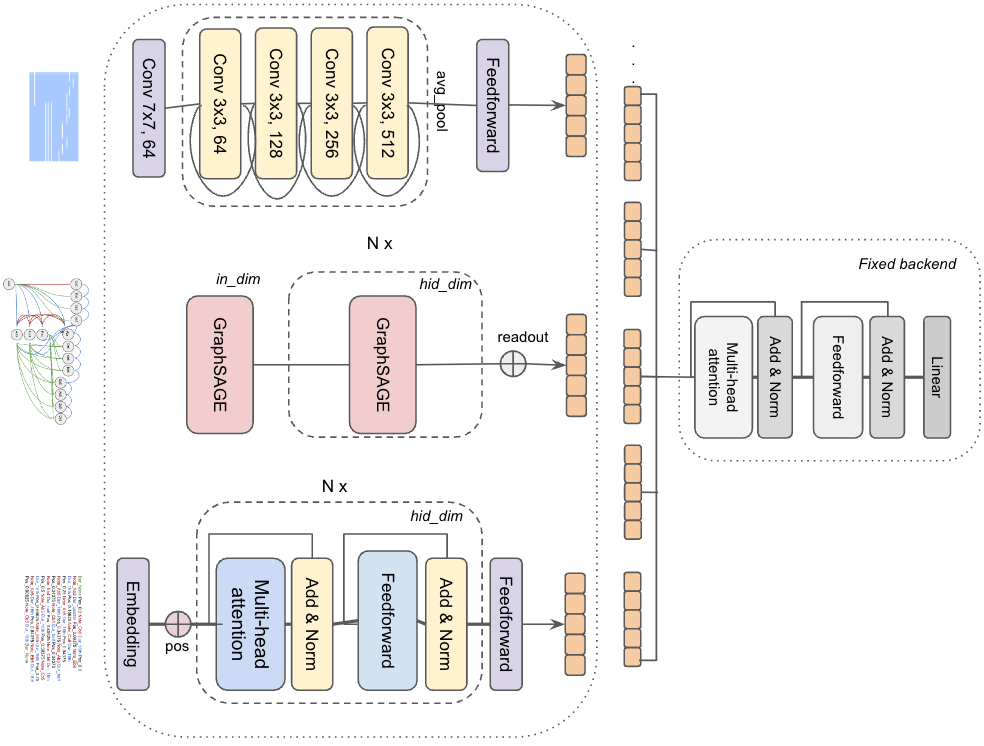}
    \caption{Left: front end for three representations, matrix, graph, and sequence, from top to bottom. Right: fixed back end with attention modules.}
    \label{fig:architecture}
\end{figure}


\subsubsection{Information Levels}
\label{sec:feature_leveling}

Given the differences in information captured by symbolic scores and performances (Sec.~\ref{sec:introduction}), we run experiments with separate levels of used information. For the base comparison experiments, we input the basic level of information that is present in both modalities: pitch, duration and onset. The advanced level of information for performance includes dynamics (MIDI velocity) and pedals, while for score includes the voice index (Sec.~\ref{sec:rep_design}) as well as score markings such as articulation and dynamics. The results and comparison of each level of information, also with respect to different tasks, will be discussed in Section~\ref{sec:music_info}. 

\begin{table*}
\footnotesize
\centering
\begin{tblr}{
  cell{1}{1} = {c=2,r=2}{},
  cell{1}{3} = {c=2}{c},
  cell{1}{5} = {c=2}{c},
  cell{1}{7} = {c=2}{c},
  cell{1}{9} = {c=2}{c},
  cell{2}{3} = {c},
  cell{2}{5} = {c},
  cell{2}{6} = {c},
  cell{2}{7} = {c},
  cell{2}{8} = {c},
  cell{2}{9} = {c},
  cell{2}{10} = {c},
  cell{3}{1} = {c=2}{c},
  cell{9}{1} = {c=2}{c},
  cell{16}{1} = {c=2}{c},
  hline{1} = {1-10}{},
  hline{3} = {1-10}{},
  hline{5,9,11,16,18,21} = {-}{},
}
                         &              & \textbf{ASAP-performance} &                     & \textbf{ASAP-score} &                     & \textbf{ATEPP-performance} &                     & \textbf{ATEPP-score} &                     \\
                         &              & \textit{ACC}              & \textit{F1}         & \textit{ACC}        & \textit{F1}         & \textit{ACC}               & \textit{F1}         & \textit{ACC}         & \textit{F1}         \\
[-1ex]\textbf{Matrix } &              &                           &                     &                     &                     &                            &                     &                      &                     \\
[-1ex]Resl             & Chnl         &                           &                     &                     &                     &                            &                     &                      &                     \\
[-1ex]400              & On+Fm        & 0.59±0.04                & 0.18±0.02          & 0.59±0.03          & 0.18±0.01          & 0.24±0.05                 & 0.20±0.04          & \textbf{0.25±0.02}  & 0.16±0.03          \\
600                      & On+Fm        & 0.62±0.06                & 0.21±0.03          & \textbf{0.61±0.07} & \textbf{0.19±0.02} & 0.28±0.01                 & \textbf{0.22±0.03} & 0.24±0.02           & 0.16±0.04          \\
800                      & Fm           & 0.62±0.04                & \textbf{0.21±0.02} & 0.58±0.06          & 0.18±0.03          & 0.22±0.03                 & 0.17±0.01          & 0.22±0.02           & \textbf{0.18±0.03} \\
800                      & On+Fm        & \textbf{0.63±0.04}       & 0.20±0.01          & 0.57±0.04          & 0.18±0.03          & \textbf{0.28±0.02}        & 0.22±0.01          & 0.22±0.04           & 0.14±0.02          \\
\textbf{Sequence }       &              &                           &                     &                     &                     &                            &                     &                      &                     \\
[-1ex]Tokn             & BPE          &                           &                     &                     &                     &                            &                     &                      &                     \\
[-1ex]MidiLike         & $\times$     & \textbf{0.53±0.05}       & \textbf{0.16±0.02} & N/A                 & N/A                 & 0.18±0.04                 & 0.10±0.02          & N/A                  & N/A                 \\
REMI                     & $\times$     & 0.51±0.04                & 0.15±0.02          & 0.43±0.04          & \textbf{0.14±0.01} & \textbf{0.23±0.04}                 & 0.10±0.02          & 0.23±0.04           & \textbf{0.13±0.02} \\
CP                       & $\times$     & 0.48±0.02                & 0.09±0.05          & \textbf{0.45±0.05} & 0.10±0.01          & 0.11±0.02                 & 0.09±0.01          & 0.17±0.06           & 0.11±0.04          \\
MidiLike                 & 4            & 0.52±0.04                & 0.15±0.02          & N/A                 & N/A                 & 0.17±0.03                 & 0.12±0.01          & N/A                  & N/A                 \\
REMI                     & 4            & 0.51±0.02                & 0.15±0.01          & 0.43±0.03          & 0.13±0.01          & 0.21±0.01                 & \textbf{0.13±0.03}          & \textbf{0.23±0.03}  & 0.13±0.01          \\
\textbf{Graph }          &              &                           &                     &                     &                     &                            &                     &                      &                     \\
[-1ex]Bi-dir           & Multi-rel    &                           &                     &                     &                     &                            &                     &                      &                     \\
[-1ex]$\times$         & $\times$     & 0.56±0.01                & 0.17±0.02          & 0.51±0.05          & 0.16±0.02          & 0.22±0.02                 & 0.10±0.03          & 0.23±0.03           & 0.21±0.05          \\
$\times$                 & $\checkmark$ & 0.58±0.03                & 0.19±0.01          & \textbf{0.54±0.05} & \textbf{0.17±0.02} & \textbf{0.27±0.03}        & 0.13±0.02          & \textbf{0.29±0.10}  & 0.18±0.06          \\
$\checkmark$             & $\checkmark$ & \textbf{0.62±0.02}       & \textbf{0.21±0.01} & 0.50±0.04          & 0.17±0.01          & 0.23±0.04                 & \textbf{0.16±0.03} & 0.27±0.06           & \textbf{0.22±0.03} 
\end{tblr}
\caption{Composer classification results for all representations, on all target subsets of our datasets on the composer classification task using only basic level features.
For each subset of data, we present the accuracy score and the macro F1 score with $8$-fold cross-validation. See Section~\ref{subsec:model_performance} for explanation of the parameters. 
}
\label{tab:base_results}
\end{table*}

\subsection{Modelling Pipelines}
\label{sec:model_pipeline}

In this work, we evaluate the input representations under the same training pipeline of different piece-level classification tasks, as discussed in Section~\ref{subsec:corpora}. We split our training architecture into two parts, a front end that projects a window of musical context into a 64-dimensional embedding, and a back end that aggregates the embedding for final prediction. The front end is representation-specific while the back end rests fixed. For a fair comparison, we ensure that the same amount of musical context is given for different front ends to learn. For MIDI performances we fix a window of \SI{60}{\second}, and for symbolic scores, we choose a window of 120 beats given that \SI{120}{\bpm} is a common tempo for music.   

For the front end, we employ a commonly used architecture for each respective representation domain:

\textbf{Matrix:} Convolutional neural network based on ResNet \cite{He2016DeepRecognition} blocks with channel numbers adapted to our input. 

\textbf{Sequence:} Transformer-encoder \cite{Vaswani2017AttentionNeed} front end with positional encoding. Each layer includes multi-head attention with 16 heads followed by an Add \& Norm layer. For the combined tokens \textit{CPWord} we add separate embedding layers for each token category in the front end. 

\textbf{Graph:} Our graph convolution network (GCN) is built by stacking GraphSAGE blocks~\cite{Hamilton2017InductiveGraphs} followed by a global mean pooling layer. We experiment with both heterogeneous and homogeneous GraphSAGE. Note that a heterogeneous network has $r$ times more parameters, where $r$ is the number of distinct edge relation types.

For the fixed back end, we used a multi-head attention block with linear projection heads to the desired number of classes, as shown in Figure~\ref{fig:architecture}. To minimize the impact of model capacity on our comparative discussion, we carried out an ablation study to understand the size of the architecture proportional to each kind of representation (Sec.~\ref{subsec:model_size}). 

\subsection{Tasks and Datasets}\label{subsec:corpora}


In this work, we focus on three tasks: composer classification, performer classification, and difficulty assessment. 
Each one of these tasks is a piece-level task since a label is attributed per piece.
The composer classification consists of predicting the composer of the piece. The performer classification involves the prediction of the performer among a list of predefined performers included in the data source. Finally, difficulty assessment involves the prediction of a number between 1-9, with 1 being easy and 9 being hard. The difficulty labels were assembled from Henle Music.\footnote{Henle Music difficulty labels, \url{https://www.henle.de/en/about-us/levels-of-difficulty-piano/}}

To evaluate the aforementioned tasks, we use two large-scale collections of Western classical piano music that contain corresponding symbolic scores (MusicXML files) and performances (MIDI files), ASAP (1067 performances, 245 scores) and ATEPP (11742 performances, 415 scores). Both datasets contain individual files per movement. 

For the composer classification task, we exclude the least populated composer classes for balance in experiments, resulting in 10 classes for the ASAP dataset and 9 classes for the ATEPP dataset. The performer classification task uses MIDI performances of ATEPP with 20 classes. For difficulty, given that both ASAP and ATEPP datasets focus on concert repertoire, the actual classes used range from difficulty 4-9.\footnote{The full distribution of the classes for each task is shown in the supplementary material.}  For all experiments, we use an eight-fold cross-validation evaluation where $85\%$ of our data is used for training and $15\%$ for testing in each fold. 

\subsection{Training}\label{subsec:training}

We performed hyperparameter optimization sweeps to determine the optimal learning rate and model hyperparameters. Our convergence criteria include early stopping at the 60 epoch breakpoint with the patience parameter set at $0.005$ on the validation accuracy. All our experiments are trained on a single A5000 GPU, and the best models, training logs, and the code is available in the repository.\footnote{
\url{https://github.com/anusfoil/SymRep}
}

\section{Experiments and Results}
\label{sec:results}

To evaluate the different representations we performed three experiments. Our first experiment focuses on a detailed comparison of the predictive accuracy of the three representations/architectures applied to the composer classification task, since it is the most well-understood task among the three. The second experiment studies the impact of model capacity (number of trainable parameters) per representation. Our last experiment investigates the effect of different levels of input features (see Section~\ref{sec:feature_leveling}) on the three tasks.

\subsection{Representations for Composer Classification}
\label{subsec:model_performance}

Our first experiment is a comparative analysis of the three representations on our two datasets, in the domains of both MIDI performance and MusicXML score with basic level features. For each representation group we test different configurations, i.e.\ for matrix we experiment with the channel (Chnl) and timestep resolution (Resl), for sequence we change the tokenization scheme (Tokn) and apply BPE, and for graph we investigate the effect of homogeneous or heterogeneous graphs (Multi-rel) and the addition of inverse edges (Bi-dir) (see Sec.~\ref{sec:rep_design}). In Table~\ref{tab:base_results}, we present for each data subset the accuracy score and the macro F1 score and their respective standard deviations under $8$-fold cross-validation (see Sec.~\ref{subsec:corpora}).

In terms of observations per representation,
the matrix representation results indicate no significant differences  under different experimental configurations. 
For sequence representations, the \textit{MIDILike} and \textit{REMI} tokenization schemes yield comparable performance. However, our experiments suggest that \textit{CPWord} is a more challenging representation to learn in the same setting. Concerning the BPE technique, no significant difference is observed between results with 4 times the original vocabulary and the non-BPE version. 

Our graph-based models exhibit similar performance regardless of the configuration of the graph edges. In particular, the effect of reverse edges is not significant, and homogeneous graph convolution already achieves similar results to heterogeneous graph convolutional models, which indicates that implicit structural information contained in the heterogeneous approach is not strictly necessary for piece-level classification tasks.

Overall, we observe that three representations show small performance differences in given experiments, with the matrix-CNN approach having the overall best metric across the experiment groups and sequence have the worst.



Finally, we would like to discuss the \textit{album effect}, which concerns the tendency of classification models to learn non-intended features, such as acoustic features in pieces of the same album~\cite{Flexer2017CloserClassification}. In our case, this effect concerns different performances of the same piece that may give away cues for classification. Training with the entire corpus of performance MIDI, which involves different interpretations of the same piece, yields an average accuracy of 90\% (see supplementary material), which is 30\% higher for the \texttt{ASAP-perf} group. To address this issue, we fix the splits to only contain unseen pieces in the test set, which reduced the accuracy score gap between performance and score. This issue has often been overlooked in literature \cite{Micchi2018NeuralClassification, Kong2020Large-scaleComposer} and a commonly-used dataset split is not piece-specific \cite{Kim2020DeepRepresentation}. Given the recent development of large score-performance datasets, we wish to establish a scientifically correct evaluation split taking into consideration the \textit{piece effect}.


\begin{figure}[t]
    \centering
    \includegraphics[width=\linewidth]{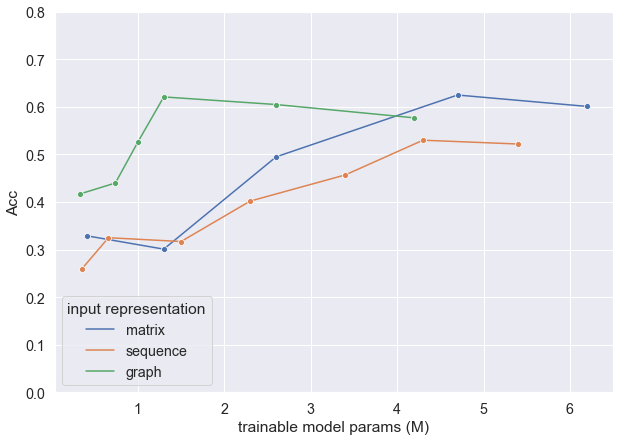}
    \caption{Model capacity vs. macro F1 score for each representation approaches on the \texttt{ASAP-composer} task. }
    \label{fig:capacity_performance}
\end{figure}

\subsection{Complexity}
\label{subsec:model_size}

In our second experiment we investigate the impact of model capacity for each representation on the composer classification task using the ASAP dataset. We experiment with different hidden dimensions $h$ and the number of layers $N$ on each architecture corresponding to each of the three representations (Sec~\ref{sec:model_pipeline}), and show our results in Figure~\ref{fig:capacity_performance}. Overall, we observe that the GCN achieves its best performance using 1.3M parameters, while architectures for matrix and sequence achieve a similar accuracy at around three times the number of parameters.

Another observation concerns the use of large models for piece-level classification tasks on symbolic data. Large convolution models such as ResNet-18/34/50 \cite{Kim2020DeepRepresentation} are substantially over-parametrized, as our results suggest we can achieve similar results using a reduced version of ResNet-8, using less than half the parameters of the smallest used ResNet architecture. Similar observations can be made for transformers, where scaling the model beyond 4.3M parameters does not further improve the performance. Our most efficient transformer encoder consists of 4 layers of attention modules with a hidden dimension of 256, significantly less than transformers used in previous related work~\cite{Fradet2023ByteMusic}. 

Finally, we note one aspect of our results after scaling our graph network. While \textit{oversmoothing} \cite{Li2019DeepGCNsCNNs} (features of graph vertices converging to the same value) is a well-known challenge to train deep GCN, our best performing model is a relatively deep and narrow network consisting of 5 layers with a hidden dimension of 64. One possible interpretation is that convergence of node features does not complicate training in the graph-level classification context.

\begin{table}
\footnotesize
\centering
\begin{tblr}{
  cell{1}{2} = {c=2}{c},
  cell{1}{4} = {c},
  cell{1}{5} = {c=2}{c},
  cell{2}{2} = {c},
  cell{2}{3} = {c},
  cell{2}{4} = {c},
  cell{2}{5} = {c},
  cell{2}{6} = {c},
  cell{3}{2} = {c},
  cell{3}{3} = {c},
  cell{3}{4} = {c},
  cell{3}{5} = {c},
  cell{3}{6} = {c},
  cell{4}{2} = {c},
  cell{4}{3} = {c},
  cell{4}{4} = {c},
  cell{4}{5} = {c},
  cell{4}{6} = {c},
  cell{5}{2} = {c},
  cell{5}{3} = {c},
  cell{5}{4} = {c},
  cell{5}{5} = {c},
  cell{5}{6} = {c},
  cell{6}{2} = {c},
  cell{6}{3} = {c},
  cell{6}{4} = {c},
  cell{6}{5} = {c},
  cell{6}{6} = {c},
  cell{7}{2} = {c},
  cell{7}{3} = {c},
  cell{7}{4} = {c},
  cell{7}{5} = {c},
  cell{7}{6} = {c},
  cell{8}{2} = {c},
  cell{8}{3} = {c},
  cell{8}{4} = {c},
  cell{8}{5} = {c},
  cell{8}{6} = {c},
  cell{9}{2} = {c},
  cell{9}{3} = {c},
  cell{9}{4} = {c},
  cell{9}{5} = {c},
  cell{9}{6} = {c},
  cell{10}{2} = {c},
  cell{10}{3} = {c},
  cell{10}{4} = {c},
  cell{10}{5} = {c},
  cell{10}{6} = {c},
  cell{11}{2} = {c},
  cell{11}{3} = {c},
  cell{11}{4} = {c},
  cell{11}{5} = {c},
  cell{11}{6} = {c},
  vline{2} = {-}{},
  hline{1,3,6,9,12} = {-}{},
}
                            & \textbf{Composer} &                & \textbf{Performer} & \textbf{Difficulty} &                \\
[-1ex]                    & perf              & score          & perf (ATEPP)       & perf                & score          \\
[-0.5ex]\textbf{Matrix}   &                   &                &                    &                     &                \\
[-1ex]basic feats~        & \textbf{0.625}    & 0.572          & \textbf{0.364}     & 0.403               & \textbf{0.420} \\
[-1ex]advanced feats      & 0.618             & \textbf{0.577}          & 0.342              & \textbf{0.411}      & 0.415          \\
[-0.5ex]\textbf{Sequence} &                   &                &                    &                     &                \\
[-1ex]basic feats         & \textbf{0.530}    & \textbf{0.447}          & 0.287              & \textbf{0.438}      & \textbf{0.368}          \\
[-1ex]advanced feats      & 0.513             & 0.393          & \textbf{0.292}              & 0.426               & 0.349          \\
[-0.5ex]\textbf{Graph}    &                   &                &                    &                     &                \\
[-1ex]basic feats         & \textbf{0.607}    & 0.545          & 0.305              & \textbf{0.373}      & 0.361          \\
[-1ex]advanced feats      & 0.598             & \textbf{0.697} & \textbf{0.323}     & 0.356               & \textbf{0.405}          \\[-0.5ex]
\end{tblr}
\caption{Accuracy of three identification tasks on the ASAP dataset, with basic or higher-level features.}
\label{tab:perf_score_task}
\end{table}

\subsection{Comparison of Feature Levels and Tasks}
\label{sec:music_info}

As discussed in Section~\ref{sec:feature_leveling}, we are also interested in understanding the impact of different levels of features on the three classification tasks. With this motivation, we performed our third set of experiments, where we adopted the best configuration of models explored in experiment 1 (see Section~\ref{subsec:model_performance}). We report the accuracy results in Table~\ref{tab:perf_score_task}. 

Our results indicate that MIDI performances and MusicXML scores have similar capabilities for distinguishing composers and difficulty. Furthermore, matrix and sequence approaches exhibit better results when learning with performances compared to scores. For the difficulty classification task, in particular, all three representations achieved approximately 40\% accuracy on the 6 difficulty levels. Performer classification is more challenging since the difference lies in the timing nuances and dynamic changes instead of the pitch information, which are more prominent in our input representations. In the 20-way classification, our approaches generally achieved around 30\% accuracy. 

Our observations suggest that the addition of advanced features has a variable impact on the representations. Interestingly, the addition of advanced features does not improve the training from sequence representations in most experiments, which can possibly be explained by the increase in vocabulary size and relative sparsity of such information. Graph structures benefit from the addition of voice edges, especially in the representation of scores, where the performance boosts for both composer and difficulty classification. Notably, the \texttt{graph-score} with advanced features configuration achieved the best result in score-based composer classification, when jointly compared with Table~\ref{tab:base_results}.

\subsection{Transformer vs.~GNN: Are We Learning the Same Set of Musical Edges?}
\label{sec:learn_edges}

A transformer can be seen as a special case of Graph Neural Networks~\cite{Velickovic2023EverythingNetworks}. Assuming a fully connected graph where vertices are tokens in a sequence, we can draw parallels between a GCN and learned attention in a transformer block. 

Therefore, we examine attention weights between \texttt{NoteOn} tokens in an effort to understand how our graph representation of the score relates to the sequence-based representation. For all pairs of \texttt{NoteOn} tokens from music sequences, we output their attention values and compute the correlation with the aggregated adjacency matrix (with all musical edges constructed in Sec.~\ref{sec:rep_design}). Across the test set of ASAP composer classification on scores, there is a weak positive correlation, with Pearson's value of 0.212. 

\begin{figure}[t]
    \centering
    \includegraphics[width=0.8\linewidth]{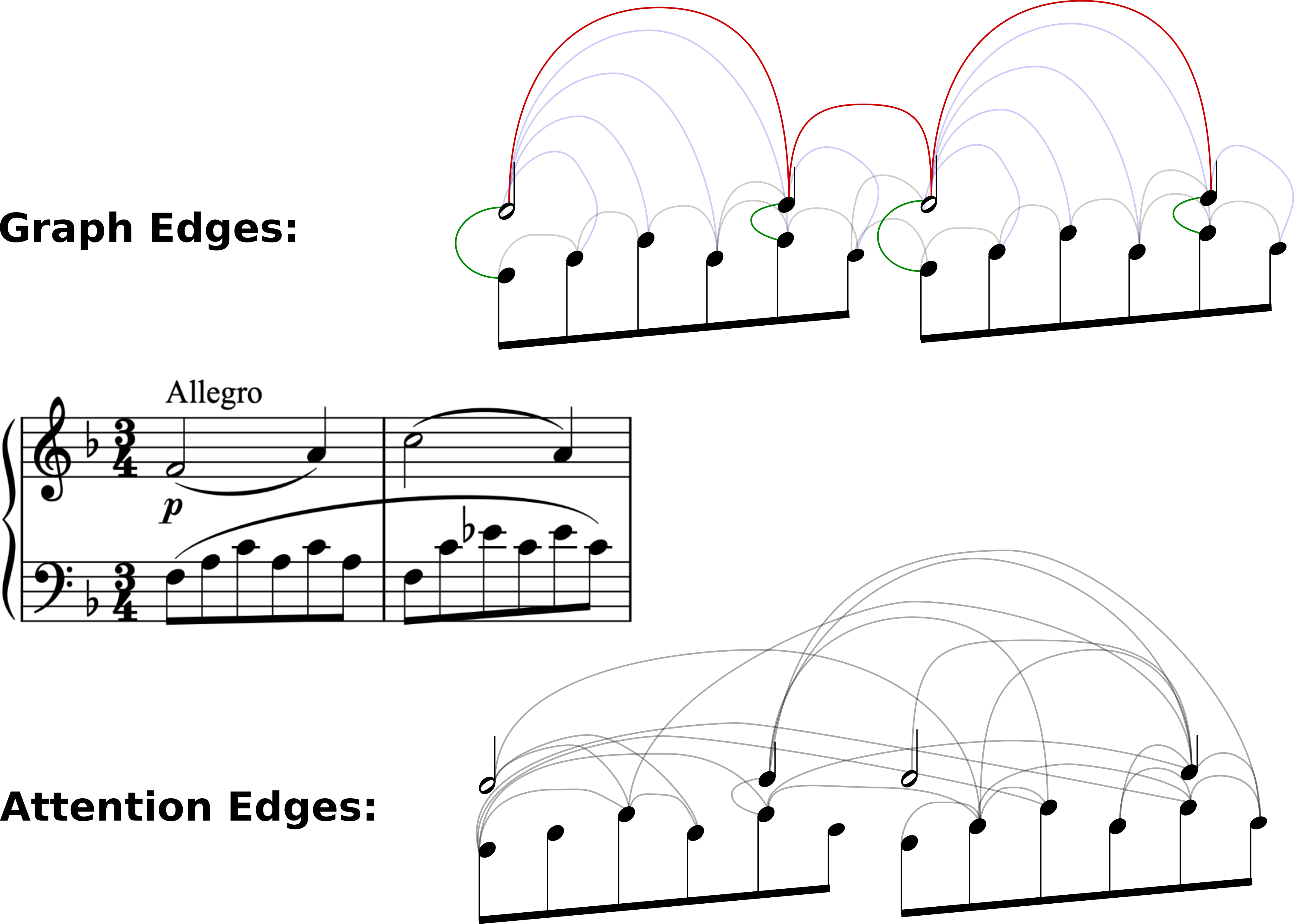}
    \caption{Visualization of graph edges (all edge types aggregated) and the attention among \texttt{NoteOn} tokens for the first measures of \textit{ Mozart Piano Sonata No.12, 1st mvt}.}
    \label{fig:attn_edge}
\end{figure}

In Figure~\ref{fig:attn_edge}, we visualize two measures of music with its constructed graph edges, and the attention across \texttt{NoteOn} tokens. We can observe some structural similarities, especially the overlap pattern in both measures, but overall the learned attention spans are much more global while graph edges connect nodes within a local range.


\section{Discussion and future work}\label{sec:conclusion}

In this paper, we presented a series of systematic experiments to investigate the impact of symbolic representations for three piece-level tasks.
In terms of simple \textit{classification performance}, we found that for a given task, different representations showed small performance differences, but no clear pattern of superiority emerged. The matrix results were marginally better on average, and usually more robust to hyper-parameter changes. More advanced features were beneficial only for certain tasks and representations.

The \textit{graph representation}, as the newest and least explored among the three approaches, exhibits promising performance, while being more light-weight (in terms of required model complexity -- cf.~Fig.~\ref{fig:capacity_performance}). We observe that homogeneous graphs produce comparable results to heterogeneous graphs for our piece-level classification tasks, and deep GCNs perform better despite over-smoothing. As graphs are arguably a more natural representation for structured artifacts such as musical scores, we believe that they should merit more detailed studies in the future.


Our model complexity experiments demonstrated that commonly used architectures in the literature are larger than necessary for our tasks, as the same results can be achieved with smaller architectures (Section \ref{subsec:model_size}).
Furthermore, we discussed the \textit{album effect} in score-performance datasets, where multiple interpretations of the same composition may cause information leakage. Our results indicate the profound impact of the album effect, and we introduce new evaluation splits to guard against this effect.

\section{Acknowledgements}
This work is supported by the UKRI Centre for Doctoral Training in Artificial Intelligence and Music, funded by UK Research and Innovation [grant number EP/S022694/1], also by the European Research Council (ERC)
under the EU's Horizon 2020 research and innovation programme, grant
agreement No.~101019375 (\textit{Whither Music?}). 


\bibliography{biblio, ref}

\clearpage

\section{Appendix}

\subsection{Album effect}

As mentioned in the main paper Section~\ref{sec:results}, the \textit{Album Effect} remains a non-trivial issue in similar classification tasks.  Here we present in Table~\ref{tab:base_results_ae} the same content as the original table (Table~\ref{tab:base_results}) from the paper which contains results from the experiment that is trained on the entire performance corpus with overlapping interpretations. Training under this non-piece-specific split, we achieved comparable accuracy (93\%) with the literature \cite{Kim2020DeepRepresentation}. 

\subsection{Complexity}

\subsubsection{Memory}

Given that the same amount of music context is input into the models, we are interested in understanding the memory efficiency of the representations. We used the native \texttt{numpy} and \texttt{cuda} functions to monitor the memory of data and memory changes during training. 

In terms the representation of a single piece of data, sequence is the most compact one while matrix takes ~70$\times$ more space, given that a lot of redundant pixels are taken in the 2D representation. The size of graph varies depending on the number of nodes and edges, but overall it is in between that of the matrix and sequence.  

However, during training we can observe that the sequence is the least memory-efficient representation during training, and it takes ~30$\times$ compares to the memory usage of matrix and graphs. Given the quadratic complexity of transformer-like architectures, the training memory needed is one of the major limitation of sequence compared to the other representations.

\begin{table}[h!]
\centering
\begin{tblr}{
  hline{2} = {-}{},
}
         & KB / seg         & KB / piece      & Training step (MB) \\
Mtr   & 819.2            & 5129.6 ± 3332.7 & 185.9 ± 105.9             \\
Seq & 12.8             & 77.8 ± 56.7     & 5548.9 ± 1736.2           \\
Gph    & 100.5 ± 57.3     & 610.9 ± 300.0   & 125.2 ± 103.4             
\end{tblr}
\label{tab:input_size}
\caption{Size estimation of each representation with basic level features from ASAP-perf data. We include the average size per segment (60s), average size per piece (as piece have different length), as well as the average allocated memory increase during each training step with a batch size of 1.}
\end{table}

\subsubsection{Convergence epochs}

During training, we also observed a difference in the time it takes the models to convergence, given the 60 epochs convergence criteria defined in Sec~\ref{subsec:training}. We first performed learning rate search using pytorch lightning's learning rate finder. Under the suggested learning rate, among different \texttt{ASAP-perf} experiment of composer classification, the matrix have on average 143.0±24.7 epochs to converge, the sequence and the graph have 132.0±31.1 and 262.0±55.7 epochs. During training, the graph models have relatively slower learning progress.

\subsection{Dataset class distributions}

We present our dataset class distribution for each task in the Table~\ref{tab:dataset_distribution}. 

\subsection{Silence and voice edges}

Besides the onset, consecutive and overlap edges in Sec~\ref{sec:rep_design}, we also add optional silence edges (edges that bridge over silence) to ensure a connected graph. A silence edge $E_{silence}$ is added between a node that's not connected by any consecutive edge and the time-wise closest node before it. The silence edge doesn't carry much music semantic meaning, and its main purpose is to prevent the disjoint subgraphs formed by distinct music sections, in which stops information flow in training. 

In the advanced representation of score graph, we input the voicing information as voice edges. Given that we can't guarantee the consistency of voice annotation in MusicXML scores (as they are mostly labeled for visual purposes like beaming), we limit the voice edge connection within a measure: If two notes are labelled with the same voice, then they are connected by a voice edge $E_{voice}$.

\begin{table*}[]
\footnotesize
\begin{tblr}{
  cell{1}{1} = {c=2,r=2}{},
  cell{1}{3} = {c=2}{c},
  cell{1}{5} = {c=2}{c},
  cell{1}{7} = {c=2}{c},
  cell{1}{9} = {c=2}{c},
  cell{2}{3} = {c},
  cell{2}{5} = {c},
  cell{2}{6} = {c},
  cell{2}{7} = {c},
  cell{2}{8} = {c},
  cell{2}{9} = {c},
  cell{2}{10} = {c},
  cell{3}{1} = {c=2}{c},
  cell{9}{1} = {c=2}{c},
  cell{16}{1} = {c=2}{c},
  hline{1,3,5,9,11,16,18,21} = {-}{},
}
                   &              & \textbf{ASAP-performance} &             & \textbf{ASAP-score} &             & \textbf{ATEPP-performance} &             & \textbf{ATEPP-score} &             \\[-1ex]
                   &              & \textit{ACC}              & \textit{F1} & \textit{ACC}        & \textit{F1} & \textit{ACC}               & \textit{F1} & \textit{ACC}         & \textit{F1} \\[-1ex]
\textbf{Matrix }   &              &                           &             &                     &             &                            &             &                      &             \\[-1ex]
Resl               & Chnl         &                           &             &                     &             &                            &             &                      &             \\[-1ex]
400                & On+Fm        & 0.926±0.02                & 0.796±0.06  & 0.598±0.03          & 0.177±0.01  & 0.905±0.04                 & 0.796±0.03  & \textbf{0.246±0.02}            &  0.156±0.03           \\
600                & On+Fm        & \textbf{0.931±0.01}                & 0.800±0.07  & \textbf{0.613±0.07}          & \textbf{0.186±0.02}  & \textbf{0.930±0.05}                 & 0.818±0.03  &  0.238±0.02                    & 0.156±0.04            \\
800                & Fm           & 0.925±0.02                & 0.723±0.11  & 0.583±0.06          & 0.182±0.03  & 0.891±0.02                 & 0.737±0.02  & 0.221±0.02           & \textbf{0.181±0.03}            \\
800                & On+Fm        & 0.926±0.02                & \textbf{0.812±0.05}  & 0.572±0.04          & 0.185±0.03  & 0.932±0.03        & \textbf{0.832±0.01}  & 0.225±0.04        &  0.138±0.02           \\
\textbf{Sequence } &              &                           &             &                     &             &                            &             &                      &             \\[-1ex]
Tokn               & BPE          &                           &             &                     &             &                            &             &                      &             \\[-1ex]
MidiLike           & $\times$     & 0.860±0.03                & 0.674±0.11  & N/A                 & N/A         & \textbf{0.926±0.01}                 & \textbf{0.769±0.01}  & N/A                  & N/A         \\
REMI               & $\times$     & 0.783±0.04                & 0.521±0.05  & 0.431±0.04          & \textbf{0.138±0.01}  & 0.910±0.01                 & 0.729±0.02  & 0.229±0.04           & \textbf{0.129±0.02}  \\
CP                 & $\times$     & 0.679±0.08                & 0.331±0.06  & \textbf{0.447±0.05}          & 0.099±0.01  & 0.864±0.02                 & 0.556±0.01  & 0.171±0.06           & 0.107±0.04  \\
MidiLike           & 4            & \textbf{0.905±0.02}                & \textbf{0.727±0.06}  & N/A                 & N/A         & 0.895±0.01                 & 0.691±0.01  & N/A                  & N/A         \\
REMI               & 4            & 0.862±0.01                & 0.692±0.07  & 0.432±0.03          & 0.132±0.01  & 0.826±0.04                 & 0.529±0.03  & \textbf{0.234±0.03}           & 0.125±0.01  \\
\textbf{Graph }    &              &                           &             &                     &             &                            &             &                      &             \\[-1ex]
Bi-dir             & Multi-rel   &                           &             &                     &             &                            &             &                      &             \\[-1ex]
$\times$           & $\times$     & 0.768±0.03                & 0.500±0.08  & 0.509±0.05          & 0.163±0.02  &  0.788±0.03                & 0.501±0.06  & 0.226±0.03           & 0.205±0.05  \\
$\times$           & $\checkmark$ & \textbf{0.861±0.03}                & \textbf{0.763±0.03}  & \textbf{0.545±0.05}          & \textbf{0.174±0.02}  & \textbf{0.928±0.01}                 & \textbf{0.781±0.03}  & \textbf{0.289±0.10}           & 0.176±0.06  \\
$\checkmark$       & $\checkmark$ & 0.833±0.03                & 0.703±0.11  & 0.500±0.04          & 0.173±0.01  & 0.897±0.01                 & 0.767±0.02  & 0.271±0.06           & \textbf{0.217±0.03}  
\end{tblr}
\caption{Base experiment composer classification results with the entire performance MIDI corpus and no piece-specific split.}
\label{tab:base_results_ae}

\end{table*}

\begin{table*}[]
\footnotesize
\centering
\begin{tblr}{
  vline{3,5,7} = {-}{},
  hline{2} = {-}{},
}
\textbf{ ASAP composer~} & \textbf{ } & \textbf{ ATEPP composer} & \textbf{ } & \textbf{ ATEPP performer } & \textbf{ } & \textbf{ ASAP difficulty } & \textbf{ } \\
Beethoven                & 195        & Beethoven                & 3033       & Richter                    & 1581       & 9                          & 164        \\
Bach                     & 163        & Chopin                   & 1739       & Ashkenazy                  & 1188       & 8                          & 176        \\
Chopin                   & 162        & Mozart                   & 653        & Arrau                      & 833        & 7                          & 132        \\
Liszt                    & 67         & Schubert                 & 264        & Brendel                    & 743        & 6                          & 150        \\
Schubert                 & 55         & Debussy                  & 254        & Kempff                     & 609        & 5                          & 56         \\
Schumann                 & 26         & Schumann                 & 243        & Barenboim                  & 603        & 4                          & 23         \\
Haydn                    & 23         & Bach                     & 231        & Schiff                     & 595        &                            &            \\
Mozart                   & 10         & Ravel                    & 169        & Horowitz                   & 576        &                            &            \\
Scriabin                 & 9          & Liszt                    & 122        & Gulda                      & 459        &                            &            \\
Ravel                    & 9          &                          &            & Gieseking                  & 362        &                            &            \\
                         &            &                          &            & Gould                      & 326        &                            &            \\
                         &            &                          &            & Gilels                     & 322        &                            &            \\
                         &            &                          &            & Perahia                    & 288        &                            &            \\
                         &            &                          &            & Pollini                    & 256        &                            &            \\
                         &            &                          &            & Argerich                   & 240        &                            &            \\
                         &            &                          &            & Schnabel                   & 240        &                            &            \\
                         &            &                          &            & François                   & 234        &                            &            \\
                         &            &                          &            & Uchida                     & 210        &                            &            \\
                         &            &                          &            & Casadesus                  & 164        &                            &            \\
                         &            &                          &            & Lugansky                   & 125        &                            &            
\end{tblr}
\label{tab:dataset_distribution}
\caption{Dataset class distribution for the tasks. The performer task is in regards to the distribution of the performed MIDI, and the other three columns are in regards to the MusicXML score.}
\end{table*}

\end{document}